\documentclass[runningheads]{llncs}
\usepackage{graphicx}
\usepackage{tikz}
\usepackage{pgfplots}
\usepackage{svg}

\begin{document}
\title{Evaluating Research Dataset Recommendations in a Living Lab}

%
\author{Jüri Keller\inst{1}\orcidID{0000-0002-9392-8646} \and
Leon Paul Mondrian Munz\inst{1}\orcidID{0000-0002-8373-5162}}
\authorrunning{J. Keller and L. Munz}
%
\institute{Technische Hochschule Köln, Ubierring 48, 50678 Cologne, Germany\\
\email{\{jueri.keller, leon\_paul\_mondrian.munz\}@smail.th-koeln.de}}
\maketitle              
\begin{abstract}
The search for research datasets is as important as laborious. Due to the importance of the choice of research data in further research, this decision must be made carefully. Additionally, because of the growing amounts of data in almost all areas, research data is already a central artifact in empirical sciences. Consequentially, research dataset recommendations can beneficially supplement scientific publication searches. We formulated the recommendation task as a retrieval problem by focussing on broad similarities between research datasets and scientific publications. In a multistage approach, initial recommendations were retrieved by the BM25 ranking function and dynamic queries. Subsequently, the initial ranking was re-ranked utilizing click feedback and document embeddings. The proposed system was evaluated live on real user interaction data using the STELLA infrastructure in the LiLAS Lab at CLEF 2021. 

Our experimental system could efficiently be fine-tuned before the live evaluation by pre-testing the system with a pseudo test collection based on prior user interaction data from the live system. The results indicate that the experimental system outperforms the other participating systems. 


\end{abstract}

\section{Introduction}
\noindent Due to the continuing flood of information and the steadily growing number of scientific publications and research datasets, the ability to find them is an ongoing challenge. Since the search for datasets, even using designated search engines, can be tedious, a possible solution may be to recommend relevant research datasets directly to corresponding publications. 

The proposed system makes use of the broad similarities between scientific publications and research datasets and is based on the probabilistic BM25 ranking function to determine the similarity between index and query~\cite{robertson1995okapi}. Results from the TREC-COVID Challenge\footnote{\url{https://ir.nist.gov/covidSubmit/index.html}} described by Roberts et al. show that almost all top-performing systems used BM25 as first stage ranker to produce already good baselines~\cite{DBLP}. By treating publications and datasets both as documents, these established retrieval techniques can be used to create dataset recommendations. The publications are used to generate queries dynamically that are subsequently used to query datasets. This initial baseline is advanced by re-ranking techniques utilizing cross-data type document embeddings and user interaction data as relevance indicators.
    
    \begin{figure}\label{website}
      \centering
      \includegraphics[width=.8\textwidth]{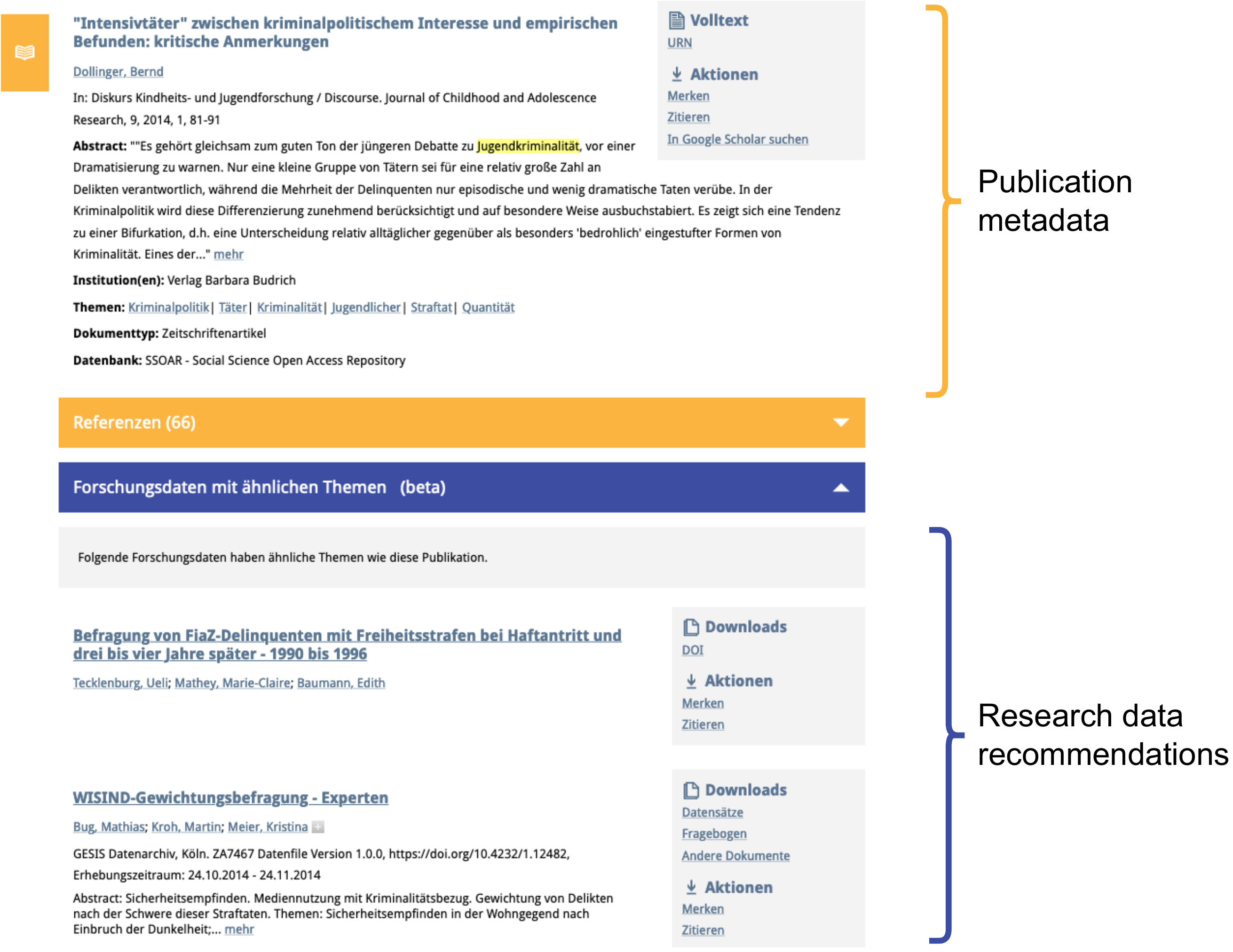}
      \caption{This figure shows an example publication detail website. Below the details of the publication, a ranking of recommended datasets is presented.}
    \end{figure}
    
The system was evaluated by the Living Labs for Academic Search (LiLAS) CLEF Challenge as a type B submission to Task 2~\cite{DBLP:conf/clef/SchaerBCWST21}. While task one focused on ad-hoc retrieval of scientific publications, Task 2 demanded for dataset recommendations related to scientific publications. Both tasks could be submitted as a pre-computed ranking (Type A) or as a live system (Type B) that retrieves the ranking ad-hoc. The STELLA infrastructure emulates pre-computed runs as live systems for queries available in the run and only utilizes the system for those queries. By that, STELLA enables the comparison of live systems with pre-computed ones~\cite{DBLP:conf/sigir/BreuerSTSWM19}. Task 2 anticipated recommender systems suggesting research datasets as a supplement to the scientific publication pages in the social science database GESIS-Search\footnote{https://search.gesis.org/}. Since all publications are known, dataset recommendations for all possible queries could be pre-computed. As a beta service solely created for the Living Lab, the website presented a ranking of a maximum of six datasets and additional metadata below the details of a publication on its overview site. The layout of the website is shown in Figure \ref{website}. More detailed task explanations and the general evaluation can be found in the lab overview~\cite{DBLP:conf/clef/SchaerBCWST21}. Living Labs differ strongly from retrieval experiments following the Cranfield paradigm and bring several unique challenges. Therefore, more authentic results can be gained. To account for the lack of initial relevance assessments and efficiently utilize the valuable user interaction feedback, we created pseudo test collections to pre-test the proposed recommender system. 

The main contribution of this work is the proposal of a BM25 based dataset recommender which is pretested with a pseudo test collection and evaluated in an online evaluation.
The remainder of this paper is structured as follows. After this introduction, the related work in the adjacent research fields is outlined in Section \ref{Related Work}. The system itself, including the pre-processing pipeline, initial ranking and re-rankings, are described in more detail in Section \ref{System}. Subsequently, the pre-testing process and process of the pseudo test collections are characterized in Sections \ref{pre testing}. After a description of results in Section \ref{results} this paper closes with a Conclusion in Section \ref{conclusion}.

\section{Related Work}\label{Related Work}
Compared to conventional Cranfield paradigm information retrieval (IR) experiments, Living Lab IR experiments aim to evaluate search systems in a real-world fashion through actual user interactions. Therefore, experimental systems extend existing search systems and are evaluated based on collected user feedback \cite{DBLP:conf/clef/SchaerBCWST21}. By that, the Living Labs for Academic Search (LiLAS) lab followed a series of labs dedicated to the living lab approach like NewsREEL \cite{DBLP:conf/mediaeval/LommatzschKHR18}, LL4IR \cite{DBLP:conf/clef/SchuthBK15} and TREC OpenSearch \cite{fd773b5b9e704bf4b4a67f52159be087} did before. Through the STELLA infrastructure the experimental systems of both types could be integrated into the live systems. Further, STELLA creates an interleaved ranking by systematically combining the results from two systems \cite{DBLP:conf/sigir/BreuerSTSWM19,DBLP:journals/dbsk/SchaibleBTMWS20}. More lifelike results and insights are expected by utilizing real user interactions to assess search systems. Azzopardi and Balog describe different stages of IR experiment environments, from the traditional test collection to the living lab \cite{DBLP:conf/clef/AzzopardiB11}.

Based on the metadata available, the dataset recommendation task was formulated as a dataset search task, recommending the retrieved datasets given a query constructed from a seed document. Chapman et al. provide a good overview of the field of dataset retrieval \cite{DBLP:journals/vldb/ChapmanSKKIKG20}. They differentiate between two types of dataset search systems, the first, most similar to document retrieval, returns existing datasets given a user query. In contrast, the second method composes a dataset based on the user query from existing data. Chapman et al. also described commonalities of datasets and documents initial retrieval systems can focus on, which may serve as a starting point in this new field \cite{DBLP:journals/vldb/ChapmanSKKIKG20}. These findings are the foundation for the proposed recommender system relying on the BM25 ranking function and a cross-datatype collection.
Kren and Mathiak analyzed dataset retrieval in the Social Sciences. They concluded that the choice of dataset is a more important and, therefore, more time-consuming decision than the choice of literature \cite{DBLP:conf/ercimdl/KernM15}. Further, even though research datasets are increasingly accessible, the connection between publications and research datasets remains often unclear~\cite{DBLP:conf/jcdl/HienertKBZM19}. Kacprzaka et al. hypothesized based on large log analyses of four open data search portals that dataset search is mainly explorative motivated \cite{DBLP:journals/ws/KacprzakKIBTS19}. This strengthens the use-case of supplementing document search results with related datasets to create a more complete overview or serve as a starting point for more exhaustive searches. The main objective should be to recommend datasets mentioned in or closely related to the seed publication.

Recommender systems are a well-discussed topic because of their broad application and the necessity to keep up with increasing information. While Bobadilla et al. \cite{Bobadilla2013RecommenderSS} provides an overview of the general field, Beel et al. \cite{Beel2015ResearchpaperRS} focus specifically on research-paper. Recommender systems use data analysis techniques to help users find the content of individual relevance. These recommender systems are often categorized into three overarching approaches based on the information source utilized to generate recommendations of, namely: Content-based recommendation, collaborative recommendation, and hybrid approaches. Content-based recommender systems primarily source the available metadata of items for recommendation. In contrast, the collaborative approaches recommend items based on user interactions with items. In conjunction of both worlds, the hybrid approaches combine collaborative, and content-based methods \cite{Adoma05}.

As initially mentioned, intermediate evaluation in a living lab setting is especially challenging because of the lacking test collection. To provide a starting point for the LiLAS lab, Schaer et al. provided head queries and candidate documents from the two real-world academic search systems, allowing the construction of pseudo test collections \cite{10.1007/978-3-030-45442-5}. Pseudo test collections are a long-established method to create synthetic queries, and relevance judgments \cite{DBLP:conf/sigir/BerendsenTWR13}. Motivated by reducing the cost of test collection creation, features are computed offline from global document information \cite{DBLP:conf/sigir/AsadiMEL11}. The provided head queries and candidates partially resemble the live system and also contain its relevance scores. Used as a pseudo test collection, they are suited to compare the experimental with the live system.

    \begin{figure}\label{schema}
      \centering
      \includegraphics[width=\textwidth]{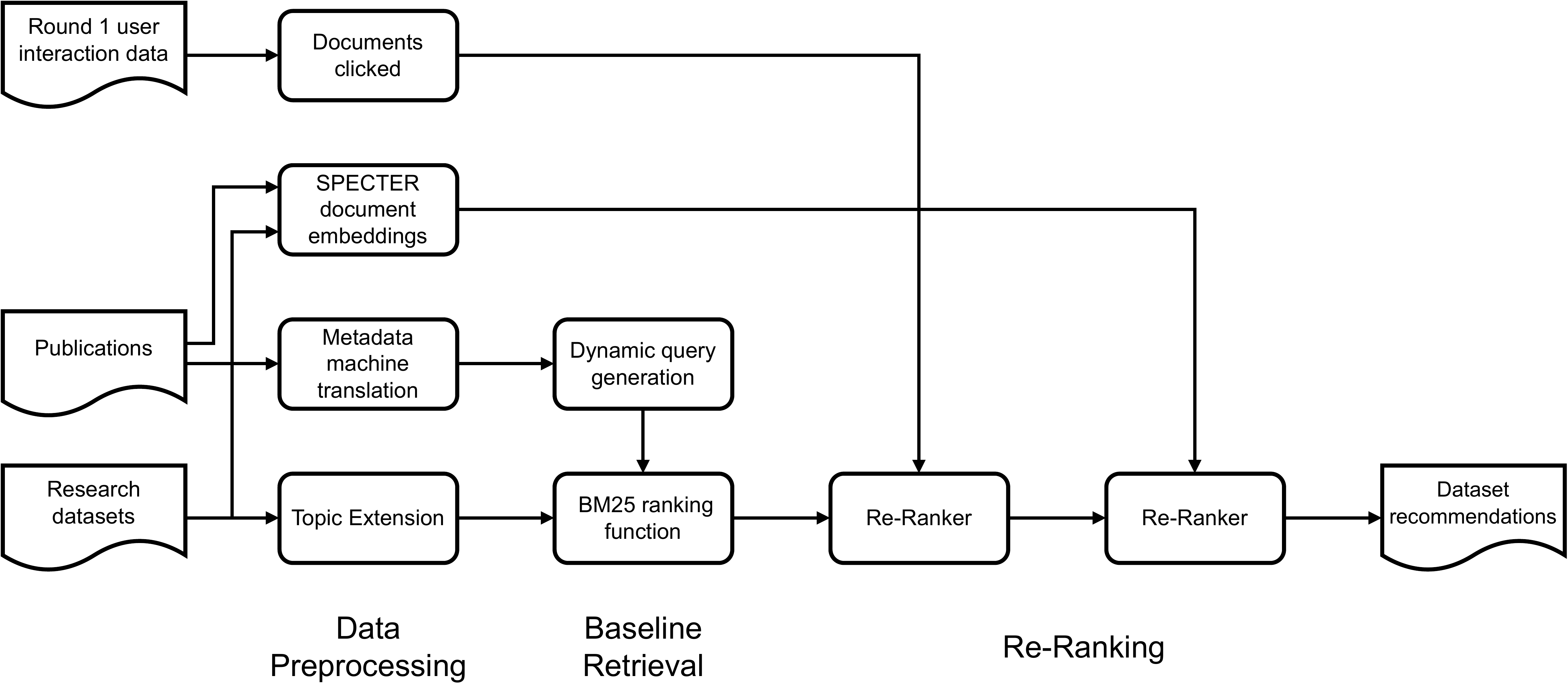}
      \caption{Visualization of the full system used to pre-compute the recommendations, from data input on the left to the final output on the right. Curvy boxes represent data inputs, rectangular boxes processing steps.}
    \end{figure}
    
\section{Research Dataset Recommendations}\label{System}
For the proposed system, the content-based approach appears to be most suitable for the prevalent use-case of recommending scientific datasets as a supplementary service during literature search. Compared to other recommendation tasks where extensive user interaction data is produced, saved in user profiles and available to fuel recommendation algorithms, ad-hoc searches, often performed without connected user profiles, provide only limited user interaction data. However, the well-established publishing practices in the scientific sector provide rich metadata for both publication and datasets. Mainly focusing on the available metadata additionally brings two advantages: First, even for niche items, in this use-case, both publications and datasets, sufficient metadata is available as a recommendation foundation. Second, while systems solely rely on user interaction data, they naturally suffer from the cold start problem where no information is available for new items; here again, the metadata of the item provides enough information. 

Since recommendations are provided in addition and related to a publication, the recommendations can be pre-computed for all available publications and need only to be updated if a new item is added. If a publication is added, new recommendations need to be calculated specifically for this item. However, all recommendations need to be updated if a dataset is added. Nevertheless, since the, admittedly by now outdated, recommendations still preserve certain relevance, they do not need to be updated right away. Therefore this approach is well suited to be pre-computed. 

The presented approach relies on the broad similarity between the provided publications and datasets, which will be described in more detail in the following subsection. By focusing on these similarities, the research dataset recommendation task is formulated as the well known and explored retrieval task. The publications research datasets need to be recommended for are used as query information, like in TREC style evaluations. By sourcing various metadata fields of a publication, a query is composed and used to retrieve the datasets which will be recommended. Initially, a baseline is retrieved using the BM25 ranking function and re-ranked by incorporating the few but available user interaction data from the first evaluation round. Since the recommendations are pre-computed but evaluated in a live environment, run time can be neglected, allowing to apply a second, more resource-intensive, neural re-ranker.

In the following, the dataset and the recommender system are described. Figure \ref{schema} gives a schematic overview of the whole process.

    \subsection{The GESIS Corpus}
    Three datasets are provided by the lab organizers originating GESIS Search for the Research Data Recommendation task: publications, datasets and candidates. 
    
    \begin{itemize}
        \item The publication dataset contains metadata for 110,420 documents from the social science database GESIS-Search\footnote{https://search.gesis.org/}. Most of the publications are provided in English or German and have textual metadata such as \textsc{title}, \textsc{abstract}, \textsc{topics} and \textsc{persons}. These publications serve as seed documents research datasets should be recommended for.
        \item In addition, metadata for 99,541 research datasets are provided. These include \textsc{title}, \textsc{topics}, \textsc{abstract}, \textsc{data type}, \textsc{collection method}, temporal and geographical coverage, \textsc{primary investigators} as well as \textsc{contributors} in English and or German. Not all metadata fields are available for all datasets.
        \item The GESIS corpus also contains collections of candidates. The top 1000 most used seed documents and their dataset recommendations are listed there. They were retrieved from the live recommender system and contain, besides the dataset identifier also, TF-IDF relevance scores.
        
    \end{itemize}
    
    Not all metadata fields were set for all datasets. To improve metadata completeness, we added machine translations of the missing title and abstract fields and systematically add missing topics.
    
    \subsection{Data Pre-processing}\label{DataEnrichment}
    Further investigations showed that the publication metadata fields for title and abstract are inconsistent in multiple ways. Not all fields are available in all languages, and not always is the actual language of a field corresponding with its label. While a German publication may have an English or partially English title and abstract, the actual language of the metadata field is of interest to correctly apply text processing to it. The publications dataset showed similar but less strong divergences. To guarantee at least one match between the related fields of a publication and a dataset, all titles and abstracts of the publications dataset are machine translated into both languages using Deep\_translator\footnote{\url{https://pypi.org/project/deep-translator/}}. 

    Not all metadata records have topics assigned. The assigned topics originate from a controlled vocabulary managed by and named after the Consortium of European Social Science Data Archives (CESSDA)\footnote{\url{https://www.cessda.eu/}}. To assign appropriate topics automatically, only existing topics are considered for assignment. A collection of all assigned topics in the corpus was created and then translated into German and English depending on their source language. To preserve the original metadata two additional fields were added: \textsc{topic\_ext\_ger} and \textsc{topic\_ext\_en}. To maintain a high topic relevance for the newly assigned topics, a topic was assigned only if it appeared in the title of the research dataset. Through this procedure, 556 German topics and 2359 English topics could be assigned.
    
    \subsection{Baseline Retrieval}
    By separating fields with multiple languages into separate fields for each language, language depending text processing could be applied to one index. The publication is used as a query to search the created index of research datasets to generate recommendations for a publication. As baseline search, Apache Solr BM25 ranking function with the default parameters $k1 = 1.2$ and $b = 0.75$ and various field combinations and boosting factors are used. Since not all fields are available for all seed publications, queries are generated dynamically considering all available fields and therefore differ in length and complexity for each publication. Each field of a seed publication is used to query the corresponding field of the research datasets. Only the topic field is queried by a concatenation of the \textsc{title}, \textsc{abstract} and \textsc{topic} fields of the seed publication. With these queries the fields \textsc{title}, \textsc{abstract} and \textsc{topic} as well as their language variations \textsc{title\_en}, \textsc{title\_de}, \textsc{abstract\_en}, \textsc{abstract\_de}, \textsc{topic\_en} and \textsc{topic\_de} and the extended topic fields \textsc{ext\_topic\_de} and \textsc{ext\_topic\_en} of the research datasets are queried if available. Each of these fields is boosted individually, considering its ability to describe the searched dataset. In general, title fields are boosted higher than abstract fields, for example. Static factors between zero and one are used as boosts to weight the scores of the fields individually.

    \subsection{Re-ranking}
    The baseline results are re-ranked in two ways to improve the recommendation quality. First, a re-ranker based on the results from round one is applied. The lab was structured in two rounds with intermediate evaluation to allow further system adjustments during the experiment. Since the described approach was only active in the second round, the results from the first round were available as additional information to re-rank the results. On top of the results re-ranked by the first re-ranker, a second re-ranker is applied, considering similarity based on document embeddings.
    
    As a signal of relevance, the click feedback from the first evaluation round is used to boost certain datasets. Given a ranking from the baseline, system datasets that were clicked in round one are boosted, considering the same query publications. Due to click sparsity and importance, a strong, static boost is added to rank the affected datasets to the top of the ranking. Incorporating user interaction data into the recommender system transforms the content-based approach into a hybrid one. This means that the recommendations need to be updated more regularly to account for ongoing variance in user interactions.
    
    Since publications and datasets have broad similarities in structure and nature, the overall document similarity is considered another factor of relevance. To measure similarity across documents and datasets, document embeddings and the k-nearest neighbors (k-NN)~\cite{fix1985discriminatory} algorithm are used. The document embeddings are calculated using SPECTER~\cite{cohan2020specter} a transformer-based SciBERT language model through its available web API\footnote{\url{https://github.com/allenai/paper-embedding-public-apis}}. From the title and abstract of a document, the language model calculates a vector that represents the document. With vectors for all documents, the documents can be mapped in a multidimensional space, and the distances between them can be measured. The closer the documents are, the bigger the similarity between them. The k-NN algorithm uses the euclidean distance to measure the distance between the documents. The closest dataset to each seed publication is calculated. Given a baseline ranking, the most similar datasets are calculated for that query publications, and all matches gain a strong static boost.

\section{System Pre-testing}\label{pre testing}
Multiple experiments were conducted to test different system configurations and determine the optimal metadata combinations and parameter settings for the field booster and re-ranker. While the predominant IR experiment type following the Cranfield paradigm relies on expensive annotated test collections, more real-world use-cases lack these amenities. These use-cases often cannot provide the required resources in terms of time and money to create a comprehensive test collection but have access to real user interaction data from live systems to evaluate experimental systems on. To maximize the efficiency of experiments and simultaneously minimize the risk of exposing potential customers to unpleasant results, it is most important to pretest the experimental systems as good as possible. Therefore, pseudo test collections may help to pretest IR experiments offline before an online evaluation with real users.

To pretest the general system and analyze the behavior of the system to specific adjustments, we created pseudo test collections from the provided head queries and candidate datasets. For each head query, the datasets recommended by the live system are provided as candidates. The pseudo test collection is constructed from all provided head queries and candidates. As most useful appeared to use the TF-IDF scores from the live system directly as relevance scores in the pseudo test collection. This pseudo test collection allowed to compare our experimental system with the TF-IDF based live system as baseline offline \cite{DBLP:conf/clef/SchaerBCWST21}.

This data holds no ground truth but can help put the results in context. Following the premise of the Living Lab evaluating experimental IR systems based on real user interactions, it is most likely that an existing system should be improved. However, since the flaws of the live system might not be known, the experimental system can only be evaluated with the live data. Therefore, differences between the live and experimental systems will be minimized at first to create a neutral starting point and then systematically deviate from that. In conclusion, the overall goal of pre-testing is to determine system settings, returning results not too far off from the baseline system but still providing enough variation for different results. All runs are evaluated using pytrec\_eval\footnote{\url{https://github.com/cvangysel/pytrec\_eval}}.

\begin{table}
    \centering
  \caption{Evaluation results for different system settings achieved during pre-testing based on the second pseudo test collection. The system producing the run number six, highlighted as italic, was submitted as final system. Except from the two results marked with the dagger ($\dagger$) the not re-ranked runs perform better then the re-ranked versions.}
  \label{tab:pre-testing}
  \begin{tabular}{|c|c|r|r|r|r|r|r|r|r|r|r|}
    \hline
    Run     & re-ranked   &     \shortstack{topic\\ boost}   &  \shortstack{abstract\\ boost}  & map       &   nDCG   &  P@5      & P@10           & R@10       & rel\_ret \\
    \hline 
    1           & False    &    0.5     &   1     &  0.077    & 0.281    & 0.273    & 0.241          & 0.024         & 0.434       \\
    2           & True     &0.5&1     & 0.077    & 0.280    &     0.269    & 0.239         & 0.024         & 0.434       \\
    3           & False    &0.7&1     & 0.070    & 0.266    &     0.256    & 0.224        & 0.023         & 0.411      \\
    4           & True     &0.7&1    & 0.070    & 0.266    &      0.255    & $\dagger$0.225         & $\dagger$0.023        & 0.411       \\
    5           & False     &0.3&1     & 0.082    & 0.292    &      \textbf{0.278}    & 0.249        & \textbf{0.025}         & 0.452       \\
    \textit{6}  & \textit{True} &\textit{0.3}&\textit{1}    &\textit{ 0.082}    & \textit{0.291}&  \textit{0.272}   &  \textit{0.246}       & \textbf{\textit{0.025}}         & \textit{0.452}      \\
    7           & False   &0.3&0.3      & 0.074    & 0.274    &     0.263   & 0.230         & 0.023        & 0.425       \\
    8           & True    &0.3&0.3      & 0.073    & 0.273    &    0.253    & 0.229          & 0.023        & 0.425       \\
    9           & False   &0.3&0.5      & \textbf{0.083}    & \textbf{0.293}    &    0.277    & 0.246          & \textbf{0.025}        & \textbf{0.453}       \\
    10          & True    &0.3&0.5      & 0.082    & 0.292    &    0.266    & 0.243          & \textbf{0.025}        & \textbf{0.453}       \\
  \hline
\end{tabular}
\end{table}

In early experiments the construction of the query was tested. Fields were added gradually to improve overall datasets retrieved and relevant datasets retrieved. Finally, the fields \textsc{title}, \textsc{title\_en} and \textsc{topic} from seed publications where used to generate the query. With this dynamically constructed query the recommendation datasets were retrieved using the original fields \textsc{title}, \textsc{abstract}, \textsc{topic} and the newly created field \textsc{ext\_topic\_de} additionally the English fields \textsc{title\_en}, \textsc{abstract\_en}, \textsc{ext\_topic\_en} were used. A second set of experiments were conducted to pretest the field boosting and re-ranker. Selected results from these experiments are shown in Table \ref{tab:pre-testing}.
All experiments were evaluated with and without re-ranking to evaluate both system components individually. Even runs, also indicated by the re-ranked field, are based on the same base system as their predecessor, but the initial results are re-ranked additionally.
The first six runs compare the three different boosting for the topic fields of the dataset metadata. The boosts 0.5, 0.7 and 0.3 are tested. Surprisingly, boosting the topics down to 0.3, tested in run 6, showed the best results.
In the remaining four runs, seven to ten negative boosts of different strengths are applied to the abstract field to account for the higher amount of words. In runs seven and eight, abstract fields are boosted down to 0.3, and in runs nine and ten, slightly less harsh, down to 0.5. 
Results, in general, are close to each other, as shown by runs five and nine, which are almost the same. Even though run nine performed slightly better in overall metrics like \textit{nDCG}, the \textit{P@5} for run five was slightly better. Since just a few recommendations can be provided on the document search result page, these metrics were prioritized, and the highlighted run six configuration was used for the final system \texttt{tekma\_n}. Both runs are based on the same initial ranking, but for run six the re-ranker was applied.

For the experiments shown in Table \ref{tab:pre-testing} almost all runs without re-ranking performed slightly better than their re-ranked versions. Only the metrics \textit{P@10} and \textit{R@10} from run four except this observation and are therefore marked with an dagger. Regardless of this, run number six was submitted as the final run to test the re-ranking approach in the live system and on the full dataset. This decision was strengthened by the observation that re-ranking could be applied to few datasets only during pre-testing.

\section{Experimental Evaluation}\label{results}
Schaer et al. provide a comprehensive evaluation of the different systems, including also weighted results accounting for the interleaved experiment setup where two systems merge their results into one result page ranking \cite{DBLP:conf/clef/SchaerBCWST21}. Additionally, this analysis focuses more specifically on the recommender system itself. STELLA, the infrastructure through which the Living Lab experiments are realized, provides detailed result feedback \cite{DBLP:conf/sigir/BreuerSTSWM19}. To quantify the individual performance of the systems in an interleaved experimental setting, Schuth et al. proposed a set of interleaving metrics \cite{DBLP:conf/clef/SchuthBK15}. Depending on the sum of results from one system clicked for one interleaved ranking in one session, the system wins, loses or results in a tie. In conjunction with the experimental system and the pre-testing runs described in Section \ref{pre testing}, the effectiveness of all ranking stages can be evaluated.
Over the course of six weeks, from 12. April to 24. May 2021 the system \texttt{tekma\_n} received 3097 Impressions. Compared to the other systems the described experimental system \texttt{tekma\_n} wins 42 times, the other experimental system \texttt{gesis\_rec\_pyterrier} wins 26 times and the baseline system \texttt{gesis\_rec\_pyserini} wins 51 times. 
However, the rankings are always composed of one experimental system and the baseline system, the baseline is utilized for twice as many sessions as an experimental system. Both experimental systems resulted in one tie and the system \texttt{tekma\_n} loses one more ranking compared to the other experimental system \texttt{gesis\_rec\_pyterrier}. All results are summarized in Table \ref{tab:results}.

\begin{table}
\centering
  \caption{Final results of Round 2, reproduced from Schaer et al. \cite{schaer2021overview}. The dagger symbol ($\dagger$) indicates the baseline system.}
  \label{tab:results}
  \begin{tabular}{|l|r|r|r|r|r|r|r|r|}
    \hline
    System  &   Win   & Loss    & Tie & Outcome   & Session   & Impression    & Clicks    &   CTR \\
    \hline
    gesis\_rec\_pyserini$\dagger$ & 51 & 68 & 2 & 0.43 & 3288 & 6034 & 53 &  0.0088\\
    gesis\_rec\_pyterrier & 26 & 25 & 1 & 0.51 & 1529 & 2937 & 27 & 0.0092 \\
    tekma\_n             & 42 & 26 & 1 & 0.62 & 1759 & 3097 & 45 & 0.0145\\
  \hline
\end{tabular}
\end{table}

The datasets clicked from the interleaved result page ranking are unevenly distributed over the ranking favoring the first positions. Since the recommendations are presented as a ranking, as illustrated in Figure \ref{website}, this exemplarily shows the position-bias in rankings \cite{DBLP:conf/wsdm/CraswellZTR08}. While datasets in the first position were clicked 21 times, datasets ranked lower were clicked less often. The full distribution of ranking positions documents that were clicked is shown in Figure 3. Considering just clicked recommendation lists, both systems, the baseline and the experimental, were utilized almost equally for the first ranking. The baseline system could rank 11 times first and the experimental system 10 times. Comparing all recommendation rankings, this finding amplifies slightly, resulting in 1021 by 958 in favor of the baseline system.

\begin{figure}
\centering
	\begin{tikzpicture}

\definecolor{darkgray176}{RGB}{176,176,176}
\definecolor{steelblue31119180}{RGB}{31,119,180}

\begin{axis} [ybar,
height=6cm,
tick align=outside,
tick pos=left,
align=center,
title={Distribution of datasets clicked per ranking position},
x grid style={darkgray176},
xlabel={Ranking position},
xtick style={draw=none},
xticklabel style={yshift=0.75ex},
scaled ticks=false, 
y grid style={darkgray176},
ylabel={Documents clicked},
ymin=0, ymax=23,
ytick style={color=black},
xtick={1,2,3,4,5,6},
ytick={0,5,10,15,20},]
\addplot[draw=none, fill=steelblue31119180] coordinates {
    (1,21) 
    (2,8) 
    (3,6) 
    (4,5)
    (5,2)
    (6,5)
};
\end{axis}
\end{tikzpicture}\label{tab:positionDistribution}
 	\begin{footnotesize}
 		\caption{Distribution of recommended datasets clicked per ranking position, reproduced from Schaer et al. \cite{schaer2021overview}. The distribution shows a position bias where high ranked datasets are clicked more often.}
 		\label{fig:sentences_lenghth}
 
 	\end{footnotesize}
 \end{figure}
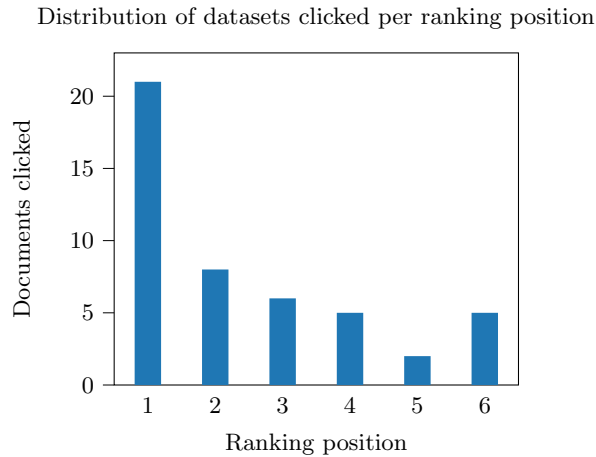

To further analyze the individual recommendations the experimental system \texttt{tekma\_n} performed worse than the comparative system, the submitted recommendations are compared with the actually clicked recommendations. The experimental system \texttt{tekma\_n} does not rank nine clicked datasets at all but ranked four datasets at the exact same position they were ranked by the baseline system and were clicked.


One main aspect of the proposed system is the data pre-processing endeavors to account for the multi-lingual data and queries described in the previous section \ref{DataEnrichment}. To measure any effects of these approaches, namely the machine translation of the title, abstract and the systematical topic expansions, the evaluated rankings are compared to rankings created during pre-testing for the same query publication. If a dataset is ranked lower without data pre-processing applied, this directly impacts being clicked for that query. Surprisingly no applied data pre-processing method, neither the translations nor the new assigned, formerly missing, topics resulted in changed positions for the clicked documents. Remembering the small basis of data, data pre-processing did not affect the results.

Following the same evaluation method, the re-ranking techniques are analyzed. Both re-rankers were assessed individually and in conjunction, but the results stayed the same. No clicked documents were re-ranked. Given these observations, the system performance observed solely relies on the query construction and initial BM25 ranking function. To achieve more comprehensive or even significant results, more user interaction is needed.

\section{Conclusion}\label{conclusion}
We applied well-established IR techniques and concepts to the fairly new field of scientific dataset recommendation and explored the early stages of IR experiments before extensive relevance assessments are available. By evaluating our endeavor in a live Living Lab experiment environment, advantages and challenges could be explored, emphasizing the differences to TREC style evaluations. By relying on real user interaction data, more authentic results can be achieved and real-world constraints can be faced. Additionally, the user interaction data provide an additional data source for the experimental system. Through extensive pre-testing based on pseudo test collections created from existing systems, the experimental system was initially aligned. Since the recommendation task could completely be pre-computed, resource extensive re-ranking techniques could be tested. By retrospectively comparing different rankings with the user click data, the impact of different ranking stages could be observed.

Results show that the applied data enrichment and re-ranking methods did not affect the position of the clicked documents. Nevertheless, our experimental system with a CTR of 0.0145 and 42 wins performed better than the other experimental system with 26 wins and a CTR of 0.0092. The results showing that the baseline only achieved 51 wins but was used in twice as many sessions indicate that our approach might outperform the baseline as well. These results must be attributed to the BM25 function and dynamic query generation. Similarities in metadata of research datasets and scientific publications allow applying these retrieval methods to create a sufficient recommender baseline. However, based on the little available user interaction data, no statistically significant results could be achieved. Nevertheless, extensive pre-testing proved to be an effective tool for achieving good results in online evaluations. Through a pseudo test collection, the recommender could be initially fine-tuned even before the online evaluation was started.

These findings can be used as reference points for future experiments at the intersection of live evaluated but pre-computed systems. Additionally, they can function as a gateway for initial systems recommending research datasets. 
In future works, multiple ranking stages could be extended. The data pre-processing could be improved to support more languages or add more topics. The user interaction data could be incorporated more distinctively by differentiating between types of interaction. This would also avoid the popularity bias, which would harm the results after a while if only clicked items are boosted. Especially interesting would be to test the system on a larger scale or longer online period to attract more user interactions as more data is required for reliable results.

\bibliographystyle{splncs04}
\bibliography{bib.bib}

\end{document}